# The Gender Balance of the Australian Space Research Community: A Snapshot from the 16th ASRC, 2016

Jonathan Horner[1], Belinda Nicholson[1], Ann Cairns[2,3,4], Wayne Short[4] and Alice Gorman[5]

[1] University of Southern Queensland, Computational Engineering and Science Research Centre, Toowoomba, Queensland 4350, Australia
[2] New South Wales Department of Education, New South Wales, Australia
[3] Division of Information Technology, Engineering and the Environment, University of South Australia, GPO Box 2471, Adelaide, South Australia, 5001, Australia
[4] National Space Society of Australia Ltd, GPO Box 7048, Sydney, New South Wales, 2001, Australia
[5] Department of Archaeology, Flinders University, GPO Box 2100, Adelaide, South Australia 5001, Australia

**Summary:** In recent years, there has been significant debate and discussion about the glaring gender disparity in the physical sciences. In order to better understand and address this issue within the Australian Space Research Community, in 2015 we began the process of keeping a statistical record of the gender balance at the annual Australian Space Research Conference. In addition, we have begun holding annual 'Women in Space Research' lunches at that conference, to discuss the situation, and to search for routes by which issues of equity can be addressed, and the situation improved.

Here, we present an update based on the 16th Australian Space Research Conference, held at RMIT, Melbourne, in late September 2016. As was the case in 2015, male attendees outnumbered female attendees by approximately 3:1. However, there was a small shift (~2.3%) in the balance, with female delegates now making up 26.4% of the total, up from 24.1% in 2015. This shift was mirrored in the gender distribution of talks, with 28.9% of all oral presentations being given by women (up from 25.2%). More striking, however, were the changes in the distribution of plenary presentations (44.4% female, up from 22.2%), poster presentations (31.8% female, up from 7.7%), and the student awards (33.3% female, up from 12.5%). These changes are encouraging, and will hopefully continue in the years to come. The conference organising committee again mirrored the gender balance of the delegates as a whole (27.3 % female vs. 26.4% of delegates), though the program committee was markedly more male-dominated this year than last (82.4% male, against last year's 72.2%).

At each year's meeting, we now hold a 'Women in Space Research' lunchtime event, where the various factors that could contribute to making the field more/less equitable are discussed. At this year's event, a number of suggestions were made that could help to make future conferences, and the wider community, a more equitable place – including increased opportunity for networking and mentoring of early- and mid-career researchers, and travel grants to make it easier for students and early-career researchers to attend the meeting. We will endeavour to put into place such schemes for future meetings, as we continue our push to make our community a more equitable place.

# Introduction

Historically, and in the present [1], the majority of science, technology, engineering and mathematical fields have been heavily male-dominated. In recent years, there has been a growing effort to better understand why this gender imbalance persists. Numerous studies indicate that *'people's behaviour is shaped by implicit or unintended biases, stemming from repeated exposure to pervasive cultural stereotypes that portray women as less competent but simultaneously emphasize their warmth and likeability compared to men'* [2].

To begin to address such biases, a number of new schemes have been established to encourage 'best practice', and reward universities that make a concerted effort to address issues of gender equity. Perhaps the most famous such scheme is the Athena SWAN (Scientific Women's Academic Network) charter, which was launched in the UK in 2005 with the goal to *'encourage and recognise commitment to advancing the careers of women in science, technology, engineering, maths and medicine (STEMM) employment in higher education and research'* [3]. The Athena SWAN charter is based on set of ten guiding principles that member institutions agree to uphold[1]. Taken together, those principles elucidate an attempt to ensure that academia is as equitable and open as possible, recognising that academia as a whole suffers wherever one part of the community is disadvantaged in comparison to another. Initially, Athena SWAN was intended to address issues of equity in the STEMM subjects, but in 2015, its remit was expanded to also cover work carried out in a number of other fields.

In Australia, the Athena SWAN model has recently been adopted by the Australian Academy of Science, who launched the Science in Gender Equity (SAGE) initiative in 2015. To date, a total of forty Australian organisations[2] have signed up to the pilot phase of the program, which aims to address the problem that *'Women comprise more than half of science PhD graduates and early career researchers, but just 17% of senior academics in Australian universities and research institutes. The loss of so many women scientists is a significant waste of expertise, talent and investment, and this impacts our nation's scientific performance and productivity.'* [4].

The global astronomical community has been particularly active in recent years in attempting to understand and address issues of gender equity. In Australia, the Astronomical Society of Australia's Inclusion, Diversity and Equity in Astronomy Chapter[3] hosts annual Diversity Workshops to build equity, and to raise awareness and understanding of a wide variety of issues that can affect the diversity of the community. In addition, in 2014 the Chapter launched the *Pleiades* awards, which are presented to those Astronomy groups across Australia which have demonstrated a commitment to equity in their workplaces[4].

A key part of the ongoing drive to make communities more equitable has been a push to gather data to allow researchers to understand the current 'state of play', and to provide a metric against which future initiatives can be assessed to see whether they are proving effective.

---

[1]   The ten principles on which the charter is based are detailed on the Athena SWAN website, at http://www.ecu.ac.uk/equality-charters/athena-swan/about-athena-swan/
[2]   The forty organisations participating in the pilot phase of the SAGE initiative are listed at https://www.sciencegenderequity.org.au/athena-swan-charter-members/
[3]   https://asa-idea.org/
[4]   https://asa-idea.org/the-pleiades-awards/

In 2014, the gender balance at the 223[rd] meeting of the American Astronomical Society was investigated in some detail in [5]. The results showed that 78 of 225 oral presentations were given by women (35%). In addition, [5] reported on the manner in which the gender of the session chair impacted upon the gender balance of the people asking questions after talks. When session chairs were male, just 20% of questions were asked by female members of the audience. With female session chairs, that number rose to 34% — a value comparable to the overall percentage of delegates that were women.

Similar results were obtained at the UK's annual National Astronomy Meeting, in 2014 [6]. At that conference, 28% of delegates were female, a gender balance that was reflected fairly in the distribution of both oral presentations and session chairs at the meeting. When it came to the questions asked at the conference, however, the authors reported the same pattern – the percentage of questions that were asked by women was markedly lower than the fraction of the overall attendees that were female.

Following these landmark studies, in 2015 we carried out an analysis of the gender balance of the 15[th] Australian Space Research Conference [7]. In brief, we found that the conference was male-dominated, with just 24% of attendees at the meeting being female. The overall gender balance of the meeting was mirrored by the distribution of oral and plenary presentations (25% and 22% female, respectively). Both the poster presentations and conference prizes were dominated by male delegates to a level greater than that expected on the basis of the distribution of attendees (male:female ratios of 12:1 and 7:1, respectively), although in both cases the number of presentations/awards was small (13 and 8)[5]. These results were intended to provide a baseline on which future conferences could be judged, to assess whether attempts to make the field more equitable bear fruit.

Here, we present the results from the 16[th] Australian Space Research Conference, held at RMIT University, in Melbourne, from the 26[th] to 28[th] September, 2016, and compare the data to that obtained in 2015. We also present a brief discussion of the key recommendations that came out of the lunchtime 'Women in Space Research' meeting, which we hope to implement in the coming years.

---

[5] To get a feel for the degree to which female delegates were under-represented in the 2015 poster and awards categories, we direct the interested reader to the online calculator at: http://aanandprasad.com/diversity-calculator/, which was brought to our attention by one of the referees of this work. That calculator shows the probability that X of Y presentations will be given by women, given the fraction of conference attendees that are female. It is based on binomial mathematics, and allows users to get a simple visualisation of the degree to which female presenters are under- or over-represented within a given sample. In the same vein, it is possible to calculate the likelihood that a given distribution of events (such as only having a single female presenter out of 13 posters) would occur by chance using such binomial mathematics. In the case of the 2015 poster presentations, such a calculation (given 24% of delegates are female) suggests that the cumulative probability of one or fewer of thirteen poster presentations being given by a women is approximately 14.4%. Similarly, one or fewer women being included in a total of eight conference awards would be expected to occur approximately 39% of the time, were those awards drawn solely by chance.

# The Gender Balance of the 16th Australian Space Research Conference

The 16th Australian Space Research Conference was a meeting with a broad scope, bringing together 174 researchers from a wide variety of disciplines. Stretching over three full days at Melbourne's RMIT University, the conference featured three parallel streams of sessions, although the whole meeting came together for the morning plenaries. As with previous years, the conference featured nine such plenaries, with speakers spanning a variety of disciplines, and career stages[6]. Conference sessions covered space engineering, space physics, cubesats, planetary science, Indigenous sky knowledge, remote sensing and GNSS (Global Navigation Satellite Systems), space business and entrepreneurship, human factors in space research, education, outreach and ethics, space situational awareness, Mars, and other space missions and projects.

As with the results we obtained in 2015, we note here that our data was obtained after the conference concluded, using the list of conference delegates. As such, the gender assignations used to build our statistics are based on our personal knowledge of the individuals concerned. The details of those individuals not personally known to us were located online, on their professional websites, and on the homepages of their institutions or employers. We used the data taken from the 2015 meeting to make this assignation process more straightforward, taking the genders assigned to repeat delegates from that dataset.

We acknowledge that this leaves open the risk that individuals might be misgendered in our dataset. Ideally, in future years, we would hope to be able to obtain this information from an anonymised survey as part of the registration process. However, it was unfortunately not possible to set up such a survey as part of the organisation for this year's meeting, and there remains some concern that delegates might not feel comfortable providing gender information as part of the registration process, even if they can be assured that such data would be anonymised and would not be linked with their abstracts, or used in any way as part of the process of constructing the conference program. This is a particularly challenging issue: on the one hand, it is clearly important to be able to track the demographics of the community as a function of time – but at the same time, we must acknowledge the role of both conscious and unconscious bias in the process by which people assess abstracts for inclusion in a conference program – and we do not wish to expose anyone in our community to discomfort around disclosing their gender. How best to obtain such information is therefore still the subject for debate amongst the organising committee, and we continue to seek advice on how best to proceed to balance these concerns.

In Table 1, we present the results obtained for the 16th Australian Space Research Conference, with the data from the 15th Australian Space Research Conference included (right hand side) for direct comparison.

It is important to note that the results presented here for the 2015 conference are slightly different to those presented in [7]. It was brought to our attention that one delegate had been misgendered as female, when in fact they identify as non-binary. We have therefore adjusted last year's tallies to take account of this – adding the column 'non-binary' to Table 1, and to Figure 1[7].

---

[6] The full program for the 16th Australian Space Research Conference can be found online at: http://www.nssa.com.au/16asrc/resources/ASRC2016-ProgramBooklet-23160923.pdf; and was last accessed on 15th December 2016 for the purposes of putting together this study.

Slightly fewer delegates attended this year's meeting than in 2015 (174 vs. 191), and whilst the number of talks was slightly lower (121 vs. 129), this was offset by an approximately commensurate increase in the number of poster presentations (22 vs. 13). Again, due to the post hoc nature of our data collection, we were unable to break down the distribution of attendees and presenters by career stage. It is possible that such data could be obtained as part of the registration process – though we note again that such data collection could raise concerns amongst those registering for the meeting that the information might be used to discriminate against them in the process of program construction. Such information therefore remains something that we would like to investigate in future years, but was not available for this work.

---

[7] Should any other delegates feel that we may have misgendered them in our assessment of either meeting, we encourage them to contact us to let us know, if they feel comfortable doing so, so that the statistics can be as accurate as possible going forward.

*Table 1: The gender distribution across the 16th Australian Space Research Conference, compared with amended numbers from the 15th Australian Space Research Conference (italicised).*

|  | 16th ASRC (2016) | | | 15th ASRC (2015) | | | |
| --- | --- | --- | --- | --- | --- | --- | --- |
|  | Male | Female | Total | *Male* | *Female* | *Non binary* | *Total* |
| Delegates | 128 73.6% | 46 26.4% | 174 | *145 75.9%* | *45 23.6%* | *1 0.5%* | *191* |
| Talks | 86 71.1% | 35 28.9% | 121 | *89 74.8%* | *30 25.2%* |  | *129* |
| Posters | 15 68.2% | 7 31.8% | 22 | *12 92.3%* | *0 0%* | *1 7.7%* | *13* |
| Plenary Presentations | 5 55.6% | 4 44.4% | 9 | *7 77.8%* | *2 22.2%* |  | *9* |
| Student Awards | 4 66.7% | 2 33.3% | 6 | *7 87.5%* | *1 12.5%* |  | *8* |
| Program Committee | 14 82.4% | 3 17.6% | 17 | *13 72.2%* | *5 27.8%* |  | *18* |
| Organising Committee | 8 72.7% | 3 27.3% | 11 | *7 70%* | *3 30%* |  | *10* |

In Figure 1, we present the gender balance of the 15th and 16th Australian Space Research Conferences, again amended as described above. From both Table 1 and Figure 1, it is clear that there was a small shift across the conference as a whole to being somewhat less male-dominated. The sole exceptions to this were the gender balances of the Program Committee and Organising Committee, which both skewed somewhat towards a more male-dominated position. This is something that we will attempt to address when organising the 17th Australian Space Research Conference, in the coming months.

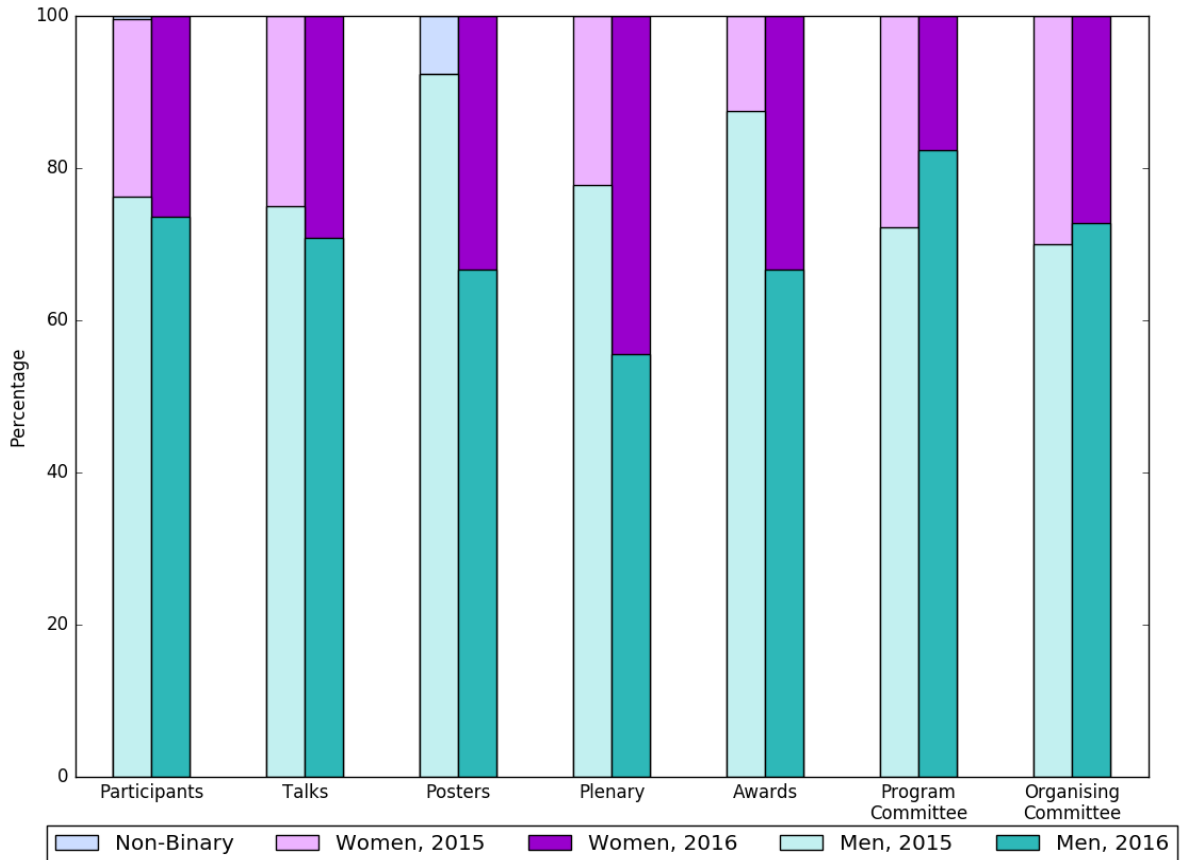

*Figure 1: The gender balance of the 15th and 16th Australian Space Research Conferences. For each category, the left-hand column shows the data from the 2015 meeting, whilst the right-hand column shows the balance for the most recent conference. It is apparent that the overall gender balance of the meeting improved in 2016 over the previous year – although we note that the program and organising committees became slightly more male-dominated.*

## The Women in Space Research Lunch

At the previous two Australian Space Research Conferences, we have held lunchtime discussion meetings in order to examine questions of equity in our community, to provide an environment where researchers can seek advice and mentoring, and to seek feedback on how the conference and our community can be made more equitable. Those meetings have been very well received, and have helped to raise awareness of issues of equity in the wider community. This series of lunchtime meetings continued at this year's ASRC, which hosted the third annual Women in Space Research Lunch.

The third annual Women in Space Research lunchtime event was once again well attended, with over forty delegates (both female and male) sitting down to discuss equity within our community. One of the key points of that lunch is that it should be a safe space[8] – and so it would be unreasonable to report specifics of the discussion in this work. We do, however, note some of the initiatives that were discussed therein, and that might be applied to our community in order to help facilitate equity, and also to benefit early- and mid-career researchers of all backgrounds.

---

[8] In other words, the lunch is intended to foster a discussion environment that is supportive and positive, and as much as possible, free from bias or criticism. The goal is that delegates should be able to freely discuss issues that they might not otherwise feel comfortable raising.

The importance of mentoring was raised, and the Astronomical Society of Australia's 'Speed Meet-A-Mentor' events (held at the Annual General Meetings of that society) was suggested as a great mechanism to help early- and mid-career researchers to connect with potential mentors from other institution, as well as building a sense of community amongst the E/MCRs themselves. It was suggested that a similar event would be welcomed as part of the 17th Australian Space Research Conference, to test the waters and see whether it could have a similarly beneficial impact on our community. Such an event could even be extended to be a 'Junior Scientists Workshop', similar to that held at the Division for Planetary Sciences meeting at Pasadena in October 2016 – perhaps taking place the afternoon before the conference, or the day after the meeting concludes.

Once again, the role of implicit bias was raised – with the 'Project Implicit' Implicit Bias tests[9] being flagged up as an excellent means by which all researchers can make themselves aware of their own inherent biases. This is particularly useful in advance of assessing grant applications, and also as part of the job-hiring process. In a similar vein, the practice of agreeing a suite of very specific assessment criteria for any given advertised position before any applications are received was recommended as both best practice as a means to avoid the impact of implicit bias, but also as a mechanism to make the whole recruitment process proceed more smoothly.

It was also suggested that there may be a number of graduate students in the community who were unable to attend the meeting, for a variety of reasons. Given that networking and meeting potential mentors are vitally important for graduate students, it was suggested that the organising committees of future meetings look at ways to make a limited number of graduate bursaries available, to try to facilitate attendance from those who would otherwise not be able.

## Progress from 2015

In [7], we described several potential initiatives that had been put forward as mechanisms by which our community could be made more equitable, with a particular focus on the annual conference itself. These recommendations were well received, and attempts were made to act upon most of them. We repeat them, below, to maintain their visibility, and follow each with an update as to the steps taken to address the initiative:

− The provision of childcare at the conference for delegates

As part of the planning of the conference, we obtained quotes for the cost of offering childcare at the conference. We were advised that at least two carers would be required, one to take care of children under the age of three, and the other to look after those who were older. We approached a number of potential sponsors to see whether it was possible to cover the costs of childcare for the meeting, but were unfortunately unsuccessful in our search. Having discussed the potential need for childcare again at the Women in Space Research lunch, it was felt that this was a relatively low priority for many of the delegates, and that our focus and funds would be better spent elsewhere (e.g. bursaries to help early-career researchers to attend).

---

9       https://implicit.harvard.edu/implicit/selectatest.html

- The option to attend the conference remotely (i.e. via streaming) for those who can't attend in person

To test the feasibility of streaming conference sessions, this year we ran live streaming for the plenary talks each day, through Zoom. The audience was limited, but we received good feedback on the quality of the streams. This remains a priority, and we will look to expand the availability of streaming at future meetings following this successful trial.

- The organisation of networking sessions during the daytime, for those who can't attend evening events

As detailed above, we again held a 'Women in Space Research' lunchtime event, which resulted in very helpful discussions, as well as offering an opportunity for networking. In future years, we intend to build upon this by running some form of 'meet-a-mentor' type event, in addition to the lunchtime meeting, to help students and early-career researchers to connect with potential mentors, and to network in a friendly and supportive environment. Mentoring is a critical activity that contributes to career success, and at least one study shows that it is routinely offered more to male students than female students [2].

- Aiming to increase the fraction of female plenary speakers above parity with the current demographic distribution of attendees

The gender balance of plenary speakers at this year's meeting was markedly improved over that at 2015 conference – with four of the nine speakers being female. This is partly the result of members of the Program Committee making deliberate choices to identify and approach women for this role. This trend will hopefully be continued at future meetings.

In the last couple of years, the phenomenon of the 'manel' – a conference session, keynote speaker line-up, or panel with no women participants – has come under increasing scrutiny in social media. It is no longer considered acceptable to have a 100% male composition of such events. This means, frequently, casting a wider net to locate women with the appropriate expertise and accepting that sometimes these women may come from outside the community.

- Inviting a plenary speaker to talk on the topic of equity in space science

Whilst none of the plenary speakers spoke on this topic, we did invite one of the plenary speakers (Professor Fran Bagenal) to speak about equity at the Women in Space Research lunch. In addition, we invited the current president of the Astronomical Society of Australia, Professor Virginia Kilborn, to talk at that meeting, discussing the efforts being made in the Astronomical community to address issues of equity, and to relate her experience in leading the push for a more equitable environment at Swinburne University. This proved a great success, and we hope that, in future, other plenary speakers with experience in this area will be willing to contribute to the lunchtime discussion meetings.

- Increasing the visibility of female space scientists online by creating Wikipedia and Scimex profiles; and creating a repository of information on women working in space science (such as pictures, CVs and biographies) to provide a teaching resource

To date, no progress has been made on these excellent suggestions. More widely, however, there is a push to address the lack of such profiles online, with events such as Hackathons and 'Wikithons' being carried out to increase the number of excellent female scientists being profiled in this manner. One such resource has been created for women in the 'trowel-wielding' professions – archaeology, geology and palaeontology, by a collective of female scholars. Their Trowelblazers website features profiles of famous and forgotten women both living and dead (http://trowelblazers.com/). The increased visibility afforded to the featured women allows others to locate them, cite their publications, and incorporate their work into teaching examples.

## Conclusions and Discussion

As part of the ongoing push to ensure that the Australian Space Research Community is as equitable as possible, we have carried out an analysis of the gender balance of the 16$^{th}$ Australian Space Research Conference, following on from our previous study of the demographics of the 15$^{th}$ such meeting, in 2015.

In 2016, there was a small improvement in the gender balance of the meeting, with an increase in the representation of women in the total pool of conference delegates (2016 was 26.4% female, as opposed to 23.6% female in 2015). This increase was mirrored by an increase in the percentage of oral presentations given by women (28.9% vs. 25.2% in 2015). The most marked improvement came in the balance of plenary presentations (4 out of 9 female in 2016, vs. 2 out of 9 in 2015), whilst the distribution of poster presentations (7 out of 22 female in 2016) and student prizes (2 of 6 female) were also improved over the distribution in 2015. Whilst these three areas all involve small numbers of participants, and so therefore are naturally subject to significant noise, it will be interesting to see whether these trends continue in future years.

Whilst the gender balance across the conference itself was closer to parity than in 2015, we note that the make-up of the organising and program committees were more male-dominated in 2016 than in 2015.

Our data support our findings from 2015 that, within the Australian Space Research community, when attending the Australian Space Research Conference, women participate as actively in presenting their research as do male delegates. The on-going problem, therefore, lies in increasing the number of women attending the conference (and addressing issues that limit female participation in the wider community).

The Australian Space Research Conference currently solicits papers in a general call out, several months before the meeting, resulting in the audience for the conference to a great extent being self-selecting. It is potentially at this stage where the current gender imbalance might be usefully addressed through schemes that enable those who feel that (for whatever reason) they are unable to attend the coming conference to have access to the meeting. For this reason, we intend to investigate further the possibility of streaming sessions (and even, potentially, allowing delegates to present remotely), as well as looking into the possibility for small travel grants being made available to those who cannot afford to otherwise attend.

It is worth noting, here, that the Australian Space Research Conference covers a wide range of sub-disciplines, within which gender disparities may be markedly different to those for the conference as a whole. To date, we have not carried out an analysis of how the gender distribution of delegates varies from discipline to discipline. Given the large number of conference sessions across the three days of the meeting, and the relatively small number of delegates who present in a given session, such an analysis may not be illuminating. Having said that, considering of the impact of the disciplines covered by the conference on its overall gender balance is clearly important.

Until fairly recently, the Australian Space Research Conference was known as the Australian Space Science Conference. The name change was adopted by the Programme Committee to attract greater participation from scholars who may not identify as STEM scientists and to reflect the existing inclusion of sessions focusing on social and cultural aspects of space science and exploration. One referee of this work suggested that this change in name may might be contributing to the small observed change in the conference's gender balance. If the gender balance of the conference is truly improving, then that should be reflected across all the disciplines covered. In future years, it would therefore be interesting to examine whether the gender balance the larger sub-disciplines (for which sufficient data can be obtained) has changed with time.

Of the 46 female and 128 male delegates at the 2016 meeting, 14 women and 43 men also attended the 2015 conference. Of the 114 delegates who attended in 2016 who did not attend the 2015 meeting, 31 were female and 83 were male. The gender balance amongst these 'new' delegates was 72.8% male: 27.2% female, reflecting the overall gender proportions at the two conferences. These figures can be used as a baseline against which to assess future retention of female conference delegates.

A number of recommendations were made in 2015 that were acted on in the organisation of the 16th Australian Space Research Conference, and it is promising that the gender balance of that meeting did improve somewhat over the previous year. However, it is important to follow these actions through for a number of years, and to monitor the demographics as a function of time, to ensure that any short-term improvements are not merely the result of noise, and to make certain that positive change in the community is propagated for future years.

## Acknowledgements

JH and BN are supported by USQ's Strategic Research Fund: the STARWINDS project, and BN is also in receipt of an Australian Postgraduate Award. The authors wish to thank the two anonymous referees of this paper for their helpful and positive feedback.

## Additional Resources

The Women in Astronomy blogspot contains a collection of excellent posts by a number of academics on a variety of topics to do with equity and gender bias. http://womeninastronomy.blogspot.com.au/

The Astronomical Society of Australia's Inclusion, Diversity and Equity in Astronomy chapter (formerly the Women In Astronomy chapter) maintains a website at https://asa-idea.org/. Of particular interest are the details of the annual Diversity Workshops, which can be found here: https://asa-idea.org/meetings/. Many of the materials from those meetings are hosted on the relevant webpages and can be freely downloaded. The IDEA chapter also hosts statistical information on the makeup of the Australian Astronomical Community, and runs the Pleiades awards, intended to recognise those astronomy departments that make a commitment to fostering an equitable environment for their staff and students.

The 'Project Implicit' Implicit Bias tests are a fascinating tool that highlights our implicit associations. These cover a wide variety of topics, from broad fields such as age, sexuality and religion, to more specifically focussed topics such as Gender-Science. They can be found here: https://implicit.harvard.edu/implicit/selectatest.html

More information on the Athena SWAN Charter can be found at http://www.ecu.ac.uk/equality-charters/athena-swan. The Science in Australia Gender Equity initiative (SAGE), which is based on the Athena SWAN charter, is detailed at https://www.sciencegenderequity.org.au/.

The Conference Diversity Distribution Calculator, http://aanandprasad.com/diversity-calculator/, is a useful tool, which allows one to quickly visualise how many women would be expected to be included in a random sample of a given size, assuming that they constitute a given percentage of the available population.